\documentclass[12pt]{article}
\usepackage{graphicx}

\newcommand\pubnumber{}
\newcommand\pubdate{\today}

\def\cea{CEA, Irfu, SPP, Centre de Saclay\\
F-91191 Gif-sur-Yvette, France}

\def\Title#1{\begin{center} {\Large #1 } \end{center}}
\def\Author#1{\begin{center}{ \sc #1} \end{center}}
\def\Address#1{\begin{center}{ \it #1} \end{center}}

\newcommand\pubblock{\rightline{\begin{tabular}{l} \pubnumber\\
         \pubdate  \end{tabular}}}
\newenvironment{Abstract}{\begin{quotation}  }{\end{quotation}}

\newenvironment{Proc}{\begin{quotation} \begin{center} \begin{large}
             Proceedings of\end{large}\end{center}\bigskip 
      \begin{center}\begin{large}}{\end{large}\end{center} \end{quotation}}

\def\babar{\mbox{\slshape B\kern-0.1em{A}\kern-0.1em
    B\kern-0.1em{A\kern-0.2em R}}}

\def\beq{\begin{equation}}
\def\eeq#1{\label{#1}\end{equation}}
\def\eeqn{\end{equation}}

\def\beqa{\begin{eqnarray}}
\def\eeqa#1{\label{#1}\end{eqnarray}}
\def\eeqan{\end{eqnarray}}

\let\bar=\overbar

\def\Dslash{\not{\hbox{\kern-4pt $D$}}}
\def\dslash{\not{\hbox{\kern-2pt $\del$}}}

\def\msb{{\bar{\ssstyle M \kern -1pt S}}}

\begin{document}
\begin{titlepage}
\pubblock

\vfill
\Title{Polarization of $B \to VV$: experimental status}
\vfill
\Author{Georges Vasseur}
\Address{\cea}
\vfill
\begin{Abstract}
The experimental status of the polarization measurements 
in $B$ to charmless vector-vector decays 
by both the Belle and \babar\ experiments is reviewed. 
The results obtained in related vector-tensor, axial vector-vector, 
and axial vector-axial vector modes are also given.
\end{Abstract}
\vfill
\begin{Proc}
CKM2010, the $6^{\rm th}$ International Workshop \\  
on the CKM Unitarity Triangle,\\
University of Warwick, UK, \\
6--10 September 2010
\end{Proc}
\vfill
\end{titlepage}
\def\thefootnote{\fnsymbol{footnote}}
\setcounter{footnote}{0}

\section{Introduction}

The amplitude for a spin zero $B$ meson decaying 
into two spin one vector mesons is the sum of three amplitudes. 
The common helicity of the two vector mesons can be 0, +1, or -1, 
corresponding to the longitudinal amplitude $A_0$ and
the tranverse amplitudes $A_{+1}$ and $A_{-1}$. 
($A_{0}$, $A_{+1}$, $A_{-1}$) defines the helicity basis.  
In the transversity basis ($A_{0}$, $A_\parallel, A_\bot$), 
with $A_\parallel = \frac{A_{+1} + A_{-1}}{\sqrt{2}}$ and 
$A_\bot = \frac{A_{+1} - A_{-1}}{\sqrt{2}}$, 
for CP eigenstates, the $A_0$ and $A_\parallel$ amplitudes are CP-even, 
while $A_\bot$ is CP-odd.

Naively, due to the V-A structure of the standard model (SM), 
$A_0$ is expected to be dominant, 
while $A_{+1}$ is suppressed by a factor $m_V/m_B$
(as well as $A_\parallel$ and $A_\bot$)
and $A_{-1}$ is further suppressed by a factor $(m_V/m_B)^2$. 
But the naive expectation is not always verified experimentally, 
in particular for penguin dominated decays. 
Several theoretical explanations have been proposed either within SM, 
such as penguin annihilation diagrams or rescattering, 
or outside SM \cite{kagan}. 

%%%%%%%%%%%%%%%%%%%%%%%%%%%%%%%%%%%%%%%%%%%%%%%%%%%%%%%%%%%%%%%%%%%%%%%%%
\begin{figure}[htb]
\centering
\includegraphics[width=100mm]{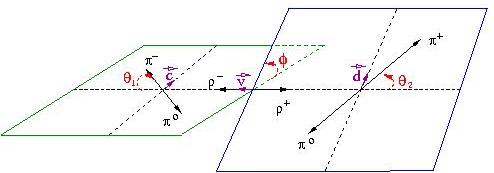}
\caption{The three observables of the angular analysis:
 $\theta_1$, $\theta_2$ and $\phi$.}
\label{fig:observables}
\end{figure}
%%%%%%%%%%%%%%%%%%%%%%%%%%%%%%%%%%%%%%%%%%%%%%%%%%%%%%%%%%%%%%%%%%%%%%%%%%%

The polarization parameters can be extracted using the angular distribution 
of the decay products of the vector mesons. 
The three physical observables are illustrated in Figure~\ref{fig:observables} 
in the case of the $B^0 \to \rho^+ \rho^-$ decay. 
The helicity angles $\theta_1$ and $\theta_2$ are the angles 
between the direction of one of the vector meson decay product and 
the opposite of the B direction in the vector meson rest frame, 
while $\phi$ is the angle between the two vector meson decay planes. 
What we want to measure are the fractions 
$f_{L,\parallel,\bot} = \frac{|A_{0,\parallel,\bot}|^2}
{|A_{0}|^2+|A_{\parallel}|^2+|A_{\bot}|^2}$ of the three partial amplitudes
(with $f_{L}+f_{\parallel}+f_{\bot} = 1$)  and the phases 
$\phi_{\parallel,\bot} = \arg{(A_{\parallel,\bot}A_{0}^*)}$ 
of each of the transverse amplitudes with respect to the longitudinal one. 
The differential decay rate is the sum of six terms.

\begin{eqnarray*}
&&\frac{8\pi}{9\Gamma} \ \frac{d^3\Gamma}{d\cos \theta_1 \ d\cos \theta_2 \ 
d\phi} = f_L \cos^2 \theta_1 \cos^2 \theta_2 + 
\frac{1}{4} \left ( 1-f_L \right ) \sin^2 \theta_1 \sin^2 \theta_2 \\
&&+ \frac{1}{4} \left ( f_{\parallel}-f_{\bot} \right ) \sin^2 \theta_1
\sin^2 \theta_2 \cos 2\phi
- \frac{1}{2} \sqrt{ f_{\bot} f_{\parallel} } 
\sin \left ( \phi_{\bot}- \phi_{\parallel} \right ) \sin^2 \theta_1
\sin^2 \theta_2 \sin 2\phi \\
&&+\frac{1}{2\sqrt{2}} \sqrt{ f_{\parallel} f_{L} } 
\cos \phi_{\parallel} 
\sin 2\theta_1 \sin 2\theta_2 \cos \phi
- \frac{1}{2\sqrt{2}} \sqrt{ f_{\bot} f_L } \sin \phi_{\bot} 
\sin 2\theta_1 \sin 2\theta_2 \sin \phi \ .
\end{eqnarray*}

The first two terms depend only on $f_L$ and 
have the usual cosine (sine) square dependence on $\theta_1$ and $\theta_2$ 
for the longitudinal (transverse) part.
Both are flat in $\phi$. 
If we integrate over $\phi$, the last four terms, 
which depend also on $f_{\parallel,\bot}$ and $\phi_{\parallel,\bot}$, 
disappear.

This leads to the two types of angular analyses that have been performed. 
Integrating over the $\phi$ angle, 
we measure only the fraction of longitudinal polarization: 
this is the partial angular analysis. 
With enough statistics, a full angular analysis can be performed 
using the three observables $\theta_1$, $\theta_2$, and also $\phi$, 
to measure not only $f_L$, but also $f_{\bot}$, $\phi_{\parallel}$, and 
$\phi_{\bot}$. 
Other parameters can also be measured such as the overall phase $\delta_0$ and 
the direct CP asymmetries between $B$ and $\bar{B}$ on these five parameters.

\section{Measurements}

\babar\ and Belle have measured $f_L$ in various charmless $B$ decays.
The experimental results with statistical and systematic errors, 
as well as the number of $B\bar{B}$ pairs used in the analyses, are summarized 
in Table~\ref{tab:fl}.

For the tree dominated $B \to \rho \rho$ decay modes,
the fraction of longitudinal polarization is found to be very close to 100\%, 
respectively 99\% in \babar\ and 94\% in Belle for $\rho^+ \rho^-$
and 95\% in $\rho^+ \rho^0$ for both experiments 
\cite{rhorho}.  
So the naive picture holds in this case. 
For the $\rho^0 \rho^0$ mode, whose tree diagram is color suppressed and 
branching ratio is a lot smaller, 
\babar\ has found a somewhat smaller value of 75\% for $f_L$ 
\cite{rhozrhoz_babar}.

\begin{figure}[htb]
\centering
\includegraphics[width=75mm]{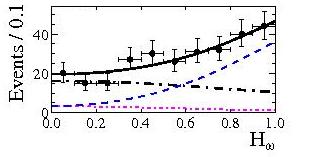}
\includegraphics[width=75mm]{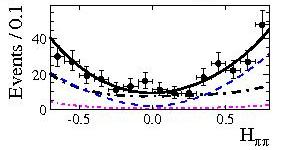}
\caption{Distribution of the $\omega$ (left) and $\rho$ (right) helicity angles
in $B^+ \to \omega \rho^+$.
Points represent data, solid curves the fit function projections, 
dashed, dotted-dashed, and long dashed-dotted curves the signal, 
$b \to c$ background, and total background.}
\label{fig:omegarho}
\end{figure}

In the $B^+ \to \omega \rho^+$ mode, which is also tree dominated,
the helicity angle distributions for the signal, 
shown in Figure~\ref{fig:omegarho},
are consistent with a dominant longitudinal polarization. 
$f_L$ is found by \babar\ to be 90\% \cite{omega}, 
close to 100\% as in $B \to \rho \rho$.

%%%%%%%%%%%%%%%%%%%%%%%%%%%%%%%%%%%%%%%%%%%%%%%%%%%%%%%%%%%%%%%%%%%%%%%%%
%%%%%%%%%%%%%%%%%%%%%%%%%%%%%%%%%%%%%%%%%%%%%%%%%%%%%%%%%%%%%%%%%%%%%%%%%%%

%%%%%%%%%%%%%%%%%%%%%%%%%%%%%%%%%%%%%%%%%%%%%%%%%%%%%%%%%%%%%%%%%%%%%%%%%%%
\begin{table}[htb]
\begin{center}
\begin{tabular}{l|cc|cc}  
Mode & \babar\ & & Belle & \\
 & $f_L$ & $N_{B\bar{B}}$ & $f_L$ & $N_{B\bar{B}}$ \\
 \hline
$B^0 \to \rho^+ \rho^-$ &
 $0.99 \pm 0.02 \,\, ^{+ \,\,0.03}_{- \,\,0.01} \,\,$ & 383M &
 $0.94 \pm 0.04 \pm 0.03$ & 535M \\
$B^+ \to \rho^+ \rho^0$ &
 $0.95 \pm 0.02 \pm 0.01$ & 465M &
 $0.95 \pm 0.11 \pm 0.02$ & \, 85M \\
$B^0 \to \rho^0 \rho^0$ &
 $0.75 \,\, ^{+\,\,0.11}_{-\,\,0.14} \,\, \pm 0.05$ & 465M &
 & \\
$B^+ \to \omega \rho^+$ &
 $0.90 \pm 0.05 \pm 0.03$ & 465M &
 & \\
 \hline
$B^0 \to \phi K^{*0}$ &
 $0.49 \pm 0.03 \pm 0.01$ & 465M &
 $0.45 \pm 0.05 \pm 0.02$ & 275M \\
$B^+ \to \phi K^{*+}$ &
 $0.49 \pm 0.05 \pm 0.03$ & 384M &
 $0.52 \pm 0.08 \pm 0.03$ & 275M \\
$B^+ \to K^{*0} \rho^+$ &
 $0.52 \pm 0.10 \pm 0.04$ & 232M &
 $0.43 \pm 0.11 \,\, ^{+ \,\,0.05}_{- \,\,0.02} \,\,$ & 275M \\
$B^0 \to K^{*0} \rho^0$ &
 $0.57 \pm 0.09 \pm 0.08$ & 232M &
 & \\
$B^0 \to \omega K^{*0}$ &
 $0.72 \pm 0.14 \pm 0.02$ & 465M &
 $0.56 \pm 0.29 \,\, ^{+ \,\,0.18}_{- \,\,0.08} \,\,$ & 657M \\
$B^+ \to \omega K^{*+}$ &
 $0.41 \pm 0.18 \pm 0.05$ & 465M &
 & \\
 \hline
$B^0 \to K^{*0} \bar{K}^{*0}$ &
 $0.80 \pm 0.11 \pm 0.06$ & 383M &
 & \\
$B^+ \to K^{*+} \bar{K}^{*0}$ &
 $0.75 \,\, ^{+ \,\,0.16}_{- \,\,0.26} \,\, \pm 0.03$ & 467M &
  & \\
 \hline
$B^0 \to \phi K_2^{*0}$ &
 $0.90 \pm 0.05 \pm 0.04$ & 465M &
 & \\
$B^+ \to \phi K_2^{*+}$ &
 $0.80 \pm 0.10 \pm 0.03$ & 465M &
 & \\
$B^0 \to \omega K_2^{*0}$ &
 $0.45 \pm 0.12 \pm 0.02$ & 465M &
 & \\
$B^+ \to \omega K_2^{*+}$ &
 $0.56 \pm 0.10 \pm 0.04$ & 465M &
 & \\
 \hline
$B^+ \to \phi K_1^{+}$ &
 $0.46 \pm 0.13 \pm 0.07$ & 465M &
 & \\
 \hline
$B^0 \to a_1^+ a_1^-$ &
 $0.31 \pm 0.22 \pm 0.10$ & 465M &
 & \\
\end{tabular}
\caption{Measurements of the fraction of longitudinal polarization $f_L$.}
\label{tab:fl}
\end{center}
\end{table}
%%%%%%%%%%%%%%%%%%%%%%%%%%%%%%%%%%%%%%%%%%%%%%%%%%%%%%%%%%%%%%%%%%%%%%%%%%%

%%%%%%%%%%%%%%%%%%%%%%%%%%%%%%%%%%%%%%%%%%%%%%%%%%%%%%%%%%%%%%%%%%%%%%%%%%%
\begin{table}[htb]
\begin{center}
\begin{tabular}{l|c|c}  
 & \babar\ & Belle \\
 \hline
$B^0 \to \phi K^{*0}$ & & \\
$f_\bot$ & $0.21 \pm 0.03 \pm 0.01$ &  $0.30 \pm 0.06 \pm 0.02$ \\
$\phi_\parallel$ & $2.40 \pm 0.13 \pm 0.08$ &  $2.39 \pm 0.24 \pm 0.04$ \\
$\phi_\bot$ & $2.35 \pm 0.13 \pm 0.09$ &  $2.51 \pm 0.23 \pm 0.04$ \\
 \hline
$B^+ \to \phi K^{*+}$ & & \\
$f_\bot$ & $0.21 \pm 0.05 \pm 0.02$ &  $0.19 \pm 0.08 \pm 0.02$ \\
$\phi_\parallel$ & $2.47 \pm 0.20 \pm 0.07$ &  $2.10 \pm 0.28 \pm 0.04$ \\
$\phi_\bot$ & $2.69 \pm 0.20 \pm 0.03$ &  $2.31 \pm 0.30 \pm 0.07$ \\
\end{tabular}
\caption{Results of the full angular analysis in $B \to \phi K^*$.}
\label{tab:full}
\end{center}
\end{table}
%%%%%%%%%%%%%%%%%%%%%%%%%%%%%%%%%%%%%%%%%%%%%%%%%%%%%%%%%%%%%%%%%%%%%%%%%%%

But for the $B \to \phi K^{*}$ penguin dominated decays, 
both experiments measure a fraction of longitudinal polarization close to 50\%:
 49\% for \babar\ and 45\% for Belle for $B^0 \to \phi K^{*0}$
and 49\% for \babar\ and 52\% for Belle for $B^+ \to \phi K^{*+}$ 
\cite{phikst}. 
This deviates from the naive expectation: 
it is referred as the polarization puzzle. 

The branching ratios in the $B \to\phi K*$ modes, about $10^{-5}$, 
are large enough.
So a full angular analysis has been performed 
by Belle and \babar\ \cite{phikst}. 
The results, shown in Table~\ref{tab:full}, 
are very similar between $B^0 \to \phi K^{*0}$ and $B^+ \to \phi K^{*+}$ modes.
The fraction of orthogonal transverse polarization is found to be about 20\%: 
21\% for \babar\ and 30\% for Belle in $B^0 \to \phi K^{*0}$ and 
21\% for \babar\ and 19\% for Belle in $B^+ \to \phi K^{*+}$. 
So $f_{\bot}$ and $f_{\parallel}$ are of the same order, 
which is consistent with the naive expectation. 
Furthermore a phase of about 2.4 radians is found 
for the two transverse amplitudes with respect to the longitudinal one 
by both experiments. 
That means that the two transverse amplitudes have a phase difference
between them  compatible with zero, 
but have non trivial phases with respect to the longitudinal amplitude.
Finally, all the direct CP parameters have been measured 
to be compatible with zero.

There are other decay modes dominated by penguin diagrams, 
such as $B \to K^* \rho$ and  $B \to \omega K^*$. 
The fraction of longitudinal polarization is measured 
by \babar\ to be 52\% for $K^{*0} \rho^+$ and 57\% for $K^{*0} \rho^0$
while Belle obtains 43\% in $K^{*0} \rho^+$ \cite{kstrho}. 
Branching ratios are smaller in $B \to \omega K^*$ 
so the precision is worse,
but the values found for $f_L$,
56\% for $\omega K^{*0}$ in Belle and 72\% for $\omega K^{*0}$ and 
41\% for $\omega K^{*+}$ in \babar\ \cite{omega} are again close to 50\%,
as for all studied decays dominated by the $b \to s$ penguin.

The $b \to d$ penguin decay, $B \to K^* K^*$,
has an even smaller branching ratio.
\babar\ measures $f_L$ to be 80\% for $K^{*0} \bar{K}^{*0}$ and 75\% for 
$K^{*+} \bar{K}^{*0}$ \cite{kstkst_babar}.  
Though the errors are large, this is an indication that
$f_L$ seems to be larger in $b \to d$ penguin than in $b \to s$ penguin:
it may be another polarization puzzle.

%%%%%%%%%%%%%%%%%%%%%%%%%%%%%%%%%%%%%%%%%%%%%%%%%%%%%%%%%%%%%%%%%%%%%%%%%
\begin{figure}[htb]
\centering
\includegraphics[width=75mm]{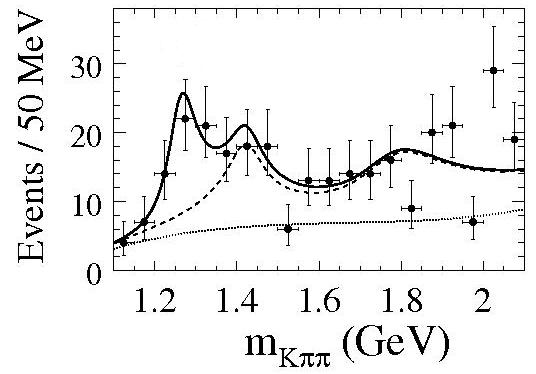}
\includegraphics[width=75mm]{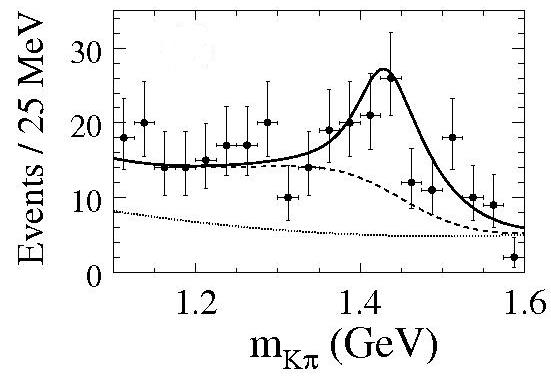}
\caption{Distribution of the $K \pi \pi$ mass (left) and $K \pi$ mass (right)
in the $B^+ \to \phi K^{*+}$ analysis.
Points represent data, 
solid curves the full fit function projections,
dotted curves the combinatorial background only, 
and dashed curves the full fit excluding the $B^0 \to \phi K_1^{0} (1270)$
signal (left) and the $B^+ \to \phi K_2^{*+} (1430)$ signal (right).}
\label{fig:highkst}
\end{figure}
%%%%%%%%%%%%%%%%%%%%%%%%%%%%%%%%%%%%%%%%%%%%%%%%%%%%%%%%%%%%%%%%%%%%%%%%%%%

Using higher $K^*$ resonances, shown in Figure~\ref{fig:highkst},
\babar\ has found a fraction of longitudinal polarization close to 50\% 
in the $B^0 \to \phi K_1^{0} (1270)$ vector-axial vector mode
\cite{phikst} and in the $B^0 \to \omega K_2^{*0} (1430)$
and $B^+ \to \omega K_2^{*+} (1430)$ vector-tensor modes \cite{omega}, 
but in contrast a large value of $f_L$
in the vector-tensor modes $B^0 \to \phi K_2^{*0} (1430)$
and $B^+ \to \phi K_2^{*+} (1430)$ \cite{phikst}.
This is another puzzle.
Finally $B^0 \to a_1^+ a_1^-$ was the first axial vector-axial vector
to be measured. 
\babar\ obtained a value of $f_L$ in the lower range \cite{a1a1_babar}.

\section{Summary}

Many vector-vector channels have been measured by Belle and \babar\ 
during the past years. 
Vector-tensor, axial vector-vector and axial vector-axial vector modes have 
also been studied. 
And a full angular analysis has been performed for the $\phi K^*$ modes. 
There are still several polarization puzzles to understand.

In summary, the fraction of longitudinal polarization is found to be large, 
closer to 100\%, for the tree dominated $\rho \rho$ and $\omega \rho$ decays, 
for the $b \to d$ penguin $K^* K^*$ modes 
and the vector-tensor $\phi K_2^{*}(1430)$ modes, 
while it is measured closer to 50\% 
for the $b \to s$ penguin dominated $\phi K^*$, $\rho K^*$, 
and $\omega K^*$ modes, 
as well as the $\omega K_2^*(1430)$ vector-tensor mode, 
the $\phi K_1(1270)$ vector-axial vector mode and 
the $a_1 a_1$ axial vector-axial vector mode.

What are the experimental perspectives? 
Belle and Babar have still further ongoing analyses. 
In the next few years, LHCb will bring new information, 
in particular for the $B_s$ to vector-vector decays. 
In the longer term a super flavor factory could study 
very rare vector-vector modes.

\end{document}